\documentclass[conference]{IEEEtran}
\usepackage{amsmath,amssymb,amsfonts,amsthm,comment,url}
\usepackage{graphicx}
\usepackage{tikz}
\usetikzlibrary{arrows.meta,positioning,fit,calc}
\usepackage{cite}
\usepackage{bm}
\newcommand{\E}{\mathbb{E}}
\usepackage{algorithm}
\usepackage[noend]{algpseudocode}

\newtheorem{lemma}{Lemma}
\newtheorem{proposition}{Proposition}
\newtheorem{remark}{Remark}

\newcommand{\bh}{\boldsymbol{h}}
\newcommand{\bff}{\boldsymbol{f}}
\newcommand{\bx}{\boldsymbol{x}}
\newcommand{\bs}{\boldsymbol{s}}
\newcommand{\by}{\boldsymbol{y}}
\newcommand{\bM}{\boldsymbol{M}}
\newcommand{\bz}{\boldsymbol{z}}
\newcommand{\bF}{\boldsymbol{F}}
\newcommand{\bJ}{\boldsymbol{J}}
\newcommand{\bg}{\boldsymbol{g}}
\newcommand{\bSigma}{\boldsymbol{\Sigma}}
\newcommand{\bn}{\boldsymbol{n}}
\newcommand{\bW}{\boldsymbol{W}}
\newcommand{\cK}{\mathcal{K}}
\newcommand{\cL}{\mathcal{L}}
\renewcommand{\E}{\mathbb{E}}
\newcommand{\NMSE}{\mathrm{NMSE}}

\begin{document}

\title{Semantic Communication via Analog RF Computing:\\ When the Radio Computes the Meaning}
\vspace{-16mm}\author{\vspace{-0mm}
\IEEEauthorblockN{Christo Kurisummoottil Thomas }
\IEEEauthorblockA{Department of Electrical and Computer Engineering, Worcester Polytechnic Institute, Worcester, MA, USA\\
Email: cthomas2@wpi.edu}\vspace{-2mm}
\vspace{-8mm}}\vspace{-8mm}

\maketitle

\begin{abstract}
Semantic communication (SC) systems extract and transmit continuous latent representations of meaning, yet the semantic encoder itself must be computed at the transmitter, and on resource-constrained edge devices this computation dominates the energy and memory budget. To overcome this challenge, in this paper, a novel end-to-end SC framework is proposed in which the semantic encoder is executed in the RF front end of the client receiver through analog RF computing. In the proposed framework, a base station (BS) broadcasts the encoder weights as RF waveforms, each edge device reuses the passive mixer in its receiver to compute its semantic latent with ultra-low energy, and the latent is then transmitted over the uplink for task inference at the network side. Analog computing noise perturbs every encoder layer while channel noise perturbs the transmitted latent, and both are shown to admit the same mathematical model. A semantic distortion metric is defined on the task posterior, and its second-order decomposition into per-client, per-stage semantic sensitivities is derived. Based on this decomposition, a closed-form semantic-aware water-filling allocation of the total client energy budget is obtained jointly over all clients, encoder layers, and uplinks, which assigns tighter accuracy targets to the stages and clients with higher semantic sensitivity. Its realization leads to a beamforming problem in which the broadcast beamformer serves the clients according to their per-layer accuracy requirements. The allocation is further refined by stochastic optimization and co-designed with multi-task noise-aware training of the shared semantic model. Simulations on MNIST show that, at matched total client energy, the unified allocation reduces the network semantic distortion by 9.2\%--19.2\% relative to uniform precision and by 5.1\%--13.1\% relative to a client-blind allocation, and noise-aware co-design reduces it by a further 46\%.
\end{abstract}

\vspace{-1mm}\section{Introduction}\vspace{-1mm}

\emph{Semantic communication (SC)} shifts the objective of a communication system from the reliable delivery of bits to the reliable delivery of meaning \cite{qin2022semantic} and \cite{Kountouris2021SemanticsEmpowered}. In the prevailing goal-oriented architecture, a learned semantic encoder at the transmitter maps a source observation into a continuous latent representation, the latent is conveyed over the channel in the style of deep joint source-channel coding (JSCC) \cite{bourtsoulatze2019deep} and \cite{shao2022learning}, and a decoder at the receiver recovers the task-relevant meaning. While a large body of work has optimized the latent transmission against channel noise, much less attention has been paid to the cost of producing the latent in the first place. On \emph{resource-constrained edge devices}, executing the semantic encoder digitally requires storing its weights and moving activations through memory, which dominates the device energy budget \cite{horowitz2014computing}. Staying within this budget requires quantizing weights and activations, which degrades task performance relative to full-precision execution.

\vspace{-2mm}\subsection{Related Works}\vspace{-2mm}
Semantic and goal-oriented communication have evolved through deep JSCC \cite{bourtsoulatze2019deep}, task-oriented compression \cite{shao2022learning}, and information-theoretic formulations \cite{gunduz2023beyond}. These works optimize the latent transmission and assume noiseless digital computation of the encoder and decoder. On the other hand, analog computing for inference has been realized with photonic processors \cite{shen2017deep} and in-memory computing \cite{ielmini2018memory}. Microwave linear analog computers perform signal processing and communication tasks with reconfigurable microwave networks at the infrastructure side \cite{nerini2025analog} and \cite{nerini2026milac}. Over-the-air computation and its neural extensions \cite{yang2020federated} and \cite{sanchez2023airnn} use the superposition property of the wireless channel as the computing medium wherein the computable functions are consequently restricted to aggregates across synchronized transmitters, and channel impairments distort the computed function itself. 

Analog radio frequency (RF) computing has recently been proposed as an inference substrate that removes the digital computing cost described above \cite{gao2026disaggregated}. A base station (BS) encodes neural network weights onto RF waveforms and broadcasts them over the downlink. Further, each client generates an input-encoded waveform locally and reuses the passive mixer in its RF front end as an analog multiplier, so that the matrix-vector multiplications (MVMs) that dominate inference are executed in the analog domain with client-side energy nearly two orders of magnitude below digital computing \cite{yu2026analogrf}. In contrast to over-the-air computation, the computation is confined to the receiver front end and is exact for arbitrary weight matrices, while channel impairments enter only as additive noise on a correct result. As a consequence, computation errors and transmission errors follow the same mathematical model, a property that over-the-air computation does not share. The physical layer design for this architecture was formulated in \cite{yu2026analogrf} from a computing-centric perspective, with beamforming and client-side scaling controlling a per-layer normalized mean squared error (NMSE), but with generic accuracy targets and no notion of transmitted meaning.

\vspace{-1mm}\subsection{Our Contributions}\vspace{-1mm}
The main contribution of this paper is an end-to-end SC architecture in which the semantic encoder is executed by analog RF computing at the client rather than stored and run digitally, and the resulting latent is transmitted over the uplink in analog JSCC fashion for decoding at the network side. This \emph{makes SC feasible on reduced-capability devices}, reduces \emph{knowledge base synchronization to changing the broadcast waveform}, and provides \emph{continuous precision control} through the analog gain of each encoder layer. However, analog computing noise at every encoder layer and channel noise on the transmitted latent both degrade the delivered meaning and draw on the same client energy budget. To the best of our knowledge, this paper is the \emph{first to close the SC loop over analog RF computing, and the first to allocate a common semantic distortion budget jointly across analog computation and wireless transmission}. 
Our key contributions include:
\begin{itemize}
\item We establish an end-to-end model of SC over analog RF computing, in which layerwise analog computing noise and uplink channel noise are unified as a single semantic noise process acting at different stages of the semantic model.
\item We define a semantic distortion metric as the expected divergence between the task posteriors produced by the clean and the perturbed pipeline, and derive a second-order decomposition into layerwise semantic sensitivities and a channel semantic sensitivity, all computable offline from a calibration dataset.
\item We obtain a closed-form semantic-aware water-filling allocation of one distortion budget jointly over all client-stage pairs, spanning the $L$ encoder layers and the uplink of every client, under a total client energy constraint. The allocation is refined by a stochastic gradient method. Its realization yields a beamforming problem in which the per-client gain requirements are determined by the semantic sensitivities of the tasks.
\item We integrate the allocation with noise-aware training of the semantic model against both noise profiles, which extends deep JSCC training to the case where the encoder computation itself is noisy. On MNIST at matched total client energy, the unified allocation reduces the network semantic distortion by $9.2\%--19.2\%$ relative to uniform precision and by $5.1\%--13.1\%$ relative to a client-blind allocation, raises fine-task accuracy from $0.51$ to $0.70$ at the tightest energy budget, and noise-aware co-design reduces distortion by a further $46\%$ at a fixed budget.
\end{itemize}

\section{System Model}

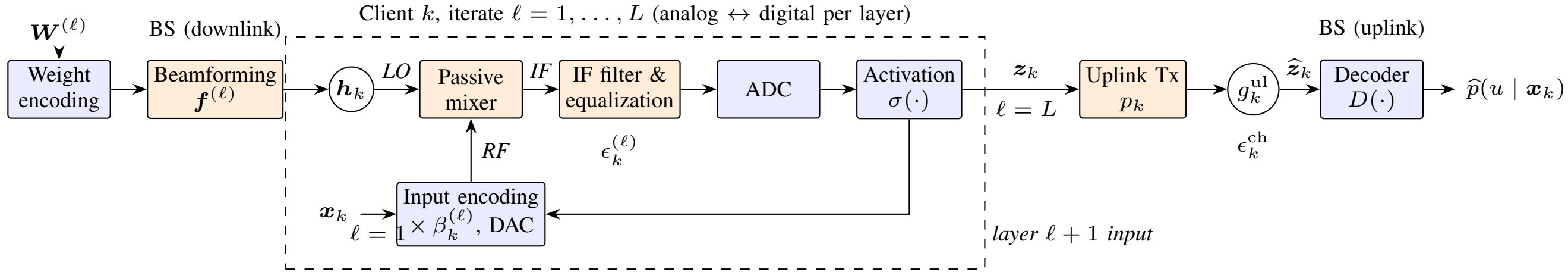
\begin{figure*}[!t]
\centering
\begin{tikzpicture}[
  font=\scriptsize,
  blk/.style={draw, rounded corners=1pt, minimum height=6.4mm, minimum width=11.4mm, align=center, inner sep=2pt},
  ana/.style={blk, fill=orange!15},
  dig/.style={blk, fill=blue!8},
  arr/.style={-{Stealth[length=1.6mm]}},
  lbl/.style={font=\scriptsize\itshape},
  node distance=3.5mm and 4mm
]
\node[dig] (wenc) {Weight\\encoding};
\node[above=1mm of wenc, font=\scriptsize] (Wl) {$\bW^{(\ell)}$};
\node[ana, right=of wenc] (bf) {Beamforming\\$\bff^{(\ell)}$};
\draw[arr] (Wl) -- (wenc);
\draw[arr] (wenc) -- (bf);
\node[right=5mm of bf, draw, circle, inner sep=1pt] (dlch) {$\bh_k$};
\draw[arr] (bf) -- (dlch);
\node[ana, right=5mm of dlch] (mix) {Passive\\mixer};
\draw[arr] (dlch) -- node[above, lbl] {LO} (mix);
\node[dig, below=7mm of mix] (ienc) {Input encoding\\$\times\,\beta_k^{(\ell)}$, DAC};
\draw[arr] (ienc) -- node[right, lbl] {RF} (mix);
\node[ana, right=of mix] (filt) {IF filter \&\\equalization};
\draw[arr] (mix) -- node[above, lbl] {IF} (filt);
\node[dig, right=of filt] (adc) {ADC};
\draw[arr] (filt) -- (adc);
\node[dig, right=of adc] (act) {Activation\\$\sigma(\cdot)$};
\draw[arr] (adc) -- (act);
\draw[arr] (act.south) |- node[below right=0mm and 8mm, lbl] {layer $\ell+1$ input} (ienc.east);
\node[left=4mm of ienc, font=\scriptsize] (xk) {$\bx_k$};
\draw[arr] (xk) -- node[below, lbl] {$\ell=1$} (ienc);
\node[below=0.5mm of filt, font=\scriptsize] {$\epsilon_k^{(\ell)}$};
\node[draw, dashed, inner sep=2.5mm, fit=(mix)(ienc)(filt)(adc)(act)(xk), label={[font=\scriptsize]above:{Client $k$, iterate $\ell = 1,\dots,L$ (analog $\leftrightarrow$ digital per layer)}}] (clientbox) {};
\node[ana, right=13mm of act] (upl) {Uplink Tx\\$p_k$};
\draw[arr] (act) -- node[above, lbl, pos=0.55] {$\bz_k$} node[below, lbl, pos=0.55] {$\ell = L$} (upl);
\node[right=4.5mm of upl, draw, circle, inner sep=1pt] (ulch) {$g_k^{\mathrm{ul}}$};
\draw[arr] (upl) -- (ulch);
\node[dig, right=4.5mm of ulch] (dec) {Decoder\\$D(\cdot)$};
\draw[arr] (ulch) -- node[above, lbl] {$\widehat{\bz}_k$} (dec);
\node[right=3.5mm of dec, font=\scriptsize] (out) {$\widehat{p}(u \mid \bx_k)$};
\draw[arr] (dec) -- (out);
\node[below=0.5mm of ulch, font=\scriptsize] {$\epsilon_k^{\mathrm{ch}}$};
\node[above=1mm of bf, font=\scriptsize] {BS (downlink)};
\node[above=1mm of dec, font=\scriptsize] {BS (uplink)};
\end{tikzpicture}
\vspace{-3mm}\caption{End-to-end SC over analog RF computing. Shaded orange blocks operate in the analog domain and shaded blue blocks in the digital domain. Here, LO: local oscillator; IF: intermediate frequency.
}
\label{fig:system}
\vspace{-4mm}\end{figure*}

We consider a multi-user multi-input multi-output (MU-MIMO) cell in which a BS with $N_t$ antennas serves $K$ single-antenna clients, indexed by $\cK \triangleq \{1,\dots,K\}$. Each client $k$ observes a local source $\bx_k$, such as an image captured by an edge camera. The network side seeks to recover a task variable $u_k$ which may indicate a class label, an event report, or a control-relevant state by decoding the transmitted latent $\boldsymbol{z}_k$. The tasks may differ across clients. The system shares one encoder $E(\cdot\,; \{\bW^{(\ell)}\}_{\ell=1}^{L})$  with $L$ linear layers, each parameterized by a weight matrix $\bW^{(\ell)} \in \mathbb{C}^{M^{(\ell)} \times N^{(\ell)}}$ ($\ell = 1,\dots,L$), interleaved with lightweight nonlinearities. The BS hosts task-specific decoders $D_k(\cdot)$, one per client or per task group. Since the encoder is shared, a single weight transmission serves all clients. The heterogeneity of the system lies in the decoders, the tasks, and the channels. At client $k$, we define
\vspace{-1mm}\begin{equation}
\bz_k = E(\bx_k), \,\, \bm{\ell}_k = D_k(\bz_k), \,\, p_k(u \mid \bx_k) = \big[\mathrm{softmax}(\bm{\ell}_k)\big]_u,
\vspace{-1mm}\end{equation}
where $\bz_k \in \mathbb{C}^{N_z}$ is the semantic latent of client $k$. The decoder outputs the logits $\bm{\ell}_k \in \mathbb{R}^{C}$ of the task posterior, where $C$ is the number of task outcomes. The encoder is not stored at the clients. It is executed through analog RF computing as follows. For each layer $\ell \in \cL \triangleq \{1,\dots,L\}$, the BS represents the entries of $\bW^{(\ell)} \in \mathbb{C}^{M^{(\ell)} \times N^{(\ell)}}$ as an RF waveform and broadcasts it after beamforming using the vector $\bff^{(\ell)} \in \mathbb{C}^{N_t}$. Client $k$ generates a local waveform representing its layer input $\bx_k^{(\ell)}$, scales it by $\beta_k^{(\ell)} \in \mathbb{C}$, and the passive mixer in its receiver multiplies this waveform with the received weight-carrying waveform to produce the layer output, following the waveform construction of \cite{yu2026analogrf,gao2026disaggregated}, as illustrated in Fig.~\ref{fig:system}. The scaling $\beta_k^{(\ell)}$ is a single complex coefficient applied to the entire input-encoded symbol, so that all entries of the layer output experience a common gain. We describe the layer accuracy using NMSE, and scalar equalization applies. The nonlinear activations between layers are lightweight elementwise operations and are applied in the client's digital baseband after filtering, equalization, and analog-to-digital conversion, following the readout chain of \cite{yu2026analogrf,gao2026disaggregated}. 
After the last encoder layer, the client maps the resulting output, which is the latent $\bz_k$ onto an analog uplink waveform and transmits it to the BS, where the decoder is executed digitally.

\vspace{-2mm}\subsection{Analog Computing Noise}\vspace{-1mm}
Let $\bh_k \in \mathbb{C}^{N_t}$ be the downlink channel of client $k$, which is assumed static over one inference pass. Further, we define  $a_k^{(\ell)} = \bh_k^{H}\bff^{(\ell)}$ as the effective gain at client $k$ of the BS beamformer $\bff^{(\ell)}$ used for layer $\ell$’s broadcast, clipped to the small-signal region of the mixer as in \cite{yu2026analogrf}. In the small-signal region, the raw mixer output for layer $\ell$ at client $k$ is $
a_k^{(\ell)}\beta_k^{(\ell)}\,\bW^{(\ell)}\bx_k^{(\ell)} + \bn_k^{(\ell)},$
with thermal noise $\bn_k^{(\ell)} \sim \mathcal{CN}(\boldsymbol{0}, \sigma_0^{2,(\ell)}\boldsymbol{I})$.  The client divides this by the estimated end-to-end gain $a_k^{(\ell)}\beta_k^{(\ell)}$ to recover \vspace{-2mm}\begin{equation}\label{eq:noisy_layer}
\widehat{\by}_k^{(\ell)} = \bW^{(\ell)} \bx_k^{(\ell)} + \widetilde{\bn}_k^{(\ell)},
\vspace{-1mm}\end{equation}
with $\widetilde{\bn}_k^{(\ell)} \sim \mathcal{CN}(\boldsymbol{0}, \sigma_{k,\mathrm{eq}}^{2,(\ell)} \boldsymbol{I})$,
where $\sigma_{k,\mathrm{eq}}^{2,(\ell)} = \sigma_0^{2,(\ell)} / \big(|a_k^{(\ell)}|^2|\beta_k^{(\ell)}|^2\big)$.
The layer inputs are generated recursively  as shown in Fig.~\ref{fig:system}, i.e.,
\begin{equation}\label{eq:recursion}
\bx_k^{(1)} = \bx_k, \qquad \bx_k^{(\ell)} = \sigma^{(\ell-1)}\big(\widehat{\by}_k^{(\ell-1)}\big), \quad \ell = 2,\dots,L,
\end{equation}
where $\sigma^{(\ell)}(\cdot)$ is the activation of layer $\ell$ applied in the digital baseband, and the latent is $\bz_k = \widehat{\by}_k^{(L)}$ for the linear final layer.  By \eqref{eq:recursion}, the input $\bx_k^{(\ell)}$ carries the propagated noise of all preceding layers, and the model \eqref{eq:noisy_layer} holds conditionally on $\bx_k^{(\ell)}$. The noise $\widetilde{\bn}_k^{(\ell)}$ in \eqref{eq:noisy_layer} is added to the pre-activation output $\bW^{(\ell)}\bx_k^{(\ell)}$, not to the post-activation input $\bx_k^{(\ell+1)}$ of the next layer. Normalizing the equalized noise power by the signal power, the pre-activation NMSE of layer $\ell$ depends on the gain only through the product magnitude \cite{yu2026analogrf},
\vspace{-2mm}\begin{equation}\label{eq:nmse}
\big(\epsilon_k^{(\ell)}\big)^2 \triangleq \NMSE_{k,\mathrm{ref}}^{(\ell)} = \frac{1}{\kappa^{(\ell)}\,|a_k^{(\ell)}|^2|\beta_k^{(\ell)}|^2},
\vspace{-2mm}\end{equation}
where $\kappa^{(\ell)} \triangleq \bar S^{(\ell)}/\sigma_0^{2,(\ell)}$ is the reference SNR of layer $\ell$, with $\bar S^{(\ell)}=\E\big[\|\bW^{(\ell)}\bx_k^{(\ell)}\|^2\big]/M^{(\ell)}$ the average clean pre-activation signal power per entry.

\vspace{-2mm}\subsection{Uplink Latent Transmission Noise}\vspace{-1mm}
We transmit the latent $\bz_k$ over the uplink by direct analog mapping of its entries onto subcarriers, in the style of analog JSCC. The uplink transmissions of different clients occupy orthogonal time-frequency resources, so no inter-client interference arises, and non-orthogonal semantic multiple access is left for future work. Let $g_k^{\mathrm{ul}}$ be the uplink channel coefficient of client $k$, assumed flat across the subcarriers carrying $\bz_k$'s entries, and $p_k$ its uplink power scaling. After matched filtering and equalization at the BS, the recovered latent is
\vspace{-2mm}\begin{equation}\label{eq:uplink}
\widehat{\bz}_k = \bz_k + \widetilde{\bn}_k^{\mathrm{ch}}, \qquad \widetilde{\bn}_k^{\mathrm{ch}} \sim \mathcal{CN}\big(\boldsymbol{0}, \sigma_{k,\mathrm{ch}}^{2} \boldsymbol{I}\big),
\vspace{-2mm}\end{equation}
where $\sigma_{k,\mathrm{ch}}^{2} = \sigma_{\mathrm{ul}}^2 / (p_k |g_k^{\mathrm{ul}}|^2)$ and $\sigma_{\mathrm{ul}}^2$ is the receiver noise power per latent entry. We define the latent root NMSE as
\vspace{-1mm}\begin{equation}\label{eq:chnmse}
\big(\epsilon_k^{\mathrm{ch}}\big)^2 \triangleq \frac{\E\big[\|\widehat{\bz}_k - \bz_k\|_2^2\big]}{\E\big[\|\bz_k\|_2^2\big]} = \frac{\sigma_{\mathrm{ul}}^2}{p_k |g_k^{\mathrm{ul}}|^2 \bar{S}_z},
\vspace{-1mm}\end{equation}
where $\bar{S}_z$ is the \emph{average latent power per entry} of the latent vector $\bz_k$, estimated from the calibration dataset and assumed common across clients since the encoder $E(\cdot)$ is shared.


We refer to the noise terms $\{\widetilde{\bn}_k^{(\ell)}\}_{\ell=1}^{L} \cup \{\widetilde{\bn}_k^{\mathrm{ch}}\}$ collectively as the \emph{semantic noise process} of the system, and collect their root NMSE values into the vector $\bm{\epsilon}_k \triangleq [\epsilon_k^{(1)},\dots,\epsilon_k^{(L)}, \epsilon_k^{\mathrm{ch}}]^{\mathsf T}$. By \eqref{eq:uplink} and \eqref{eq:noisy_layer}, the uplink channel enters as additive noise on the semantic signal in exactly the same form as an encoder layer, so a single energy budget can be allocated jointly across encoder layers and the uplink, despite their different physical origin.

\vspace{-1mm}\subsection{Client Energy Model}\vspace{-1mm}
The client's total energy consumption has two components. Fixing the mixer's beamforming gain at its maximum linear value $a_{\max}$, following \cite{yu2026analogrf}, leaves the client-side scaling $\beta_k^{(\ell)}$ as the only free variable in \eqref{eq:nmse}, so a target root NMSE $\epsilon_k^{(\ell)}$ requires
$
\big|\beta_k^{(\ell)}\big|^2 = \frac{1}{\kappa^{(\ell)} a_{\max}^2 \big(\epsilon_k^{(\ell)}\big)^2}.$
Separately, generating a waveform of scaling $\beta_k^{(\ell)}$ costs energy $c_k^{(\ell)}|\beta_k^{(\ell)}|^2$, where $c_k^{(\ell)}$ is a hardware constant \cite{yu2026analogrf} relating waveform amplitude to generation energy and does not depend on the NMSE target. Substituting the required $|\beta_k^{(\ell)}|^2$ from above, the waveform generation energy for layer $\ell$ becomes inversely proportional to $(\epsilon_k^{(\ell)})^2$. Uplink transmission of the latent at power $p_k$ over the latent transmission duration similarly costs energy inversely proportional to $(\epsilon_k^{\mathrm{ch}})^2$, by \eqref{eq:chnmse}. Substituting $\gamma_k^{(s)} \triangleq 1/(\epsilon_k^{(s)})^2$ for every stage $s \in \mathcal{S} \triangleq \{1,\dots,L,\mathrm{ch}\}$, the total client-side energy of the network is
\vspace{-3mm}\begin{equation}
E_{\mathrm{tot}} = \sum_{k \in \mathcal{K}} \sum_{s \in \mathcal{S}} \omega_{k,s}\, \gamma_k^{(s)},\label{eq:energy}
\vspace{-3mm}\end{equation}
where $\omega_{k,\ell} \triangleq \dfrac{c_k^{(\ell)}}{a_{\max}^2 \kappa^{(\ell)}}$ is the energy cost per unit of $\gamma_k^{(\ell)}$ at layer $\ell$, and $\omega_{k,\mathrm{ch}} \triangleq \dfrac{T_{\mathrm{ul}} \sigma_{\mathrm{ul}}^2}{\rho_{\mathrm{radio}} \bar S_z |g_k^{\mathrm{ul}}|^2}$ is the analogous coefficient for the uplink, with $T_{\mathrm{ul}}$ the latent transmission duration and $\rho_{\mathrm{radio}}$ the RF chain efficiency.

\vspace{-1mm}\section{Unified Semantic Distortion Allocation}\vspace{-1mm}
In this section, we define a semantic distortion metric on the task posterior, decompose it into per-client, per-stage semantic sensitivities, and use these sensitivities to allocate the client energy budget jointly over the encoder layers and the uplink of every client so as to minimize the network semantic distortion.
\vspace{-5mm}\subsection{Semantic Distortion Metric}\vspace{-1mm}
Let $p_k(u \mid \bx_k)$ be the task posterior of client $k$ under ideal channel and computation, i.e., $\bm{\epsilon}_k = \boldsymbol{0}$, and $\widehat{p}_k(u \mid \bx_k; \bm{\epsilon}_k)$ is the posterior obtained by applying the same decoder to the noisy latent. We define the \emph{semantic distortion} of client $k$ as
\vspace{-2mm}\begin{equation}\label{eq:semdist} \Delta_k(\bm{\epsilon}_k) \triangleq \E_{\bx_k,\, \{\widetilde{\bn}\}}\Big[ \mathrm{KL}\big( p_k(u \mid \bx_k) \,\|\, \widehat{p}_k(u \mid \bx_k; \bm{\epsilon}_k) \big) \Big], \vspace{-2mm}\end{equation}
where $\mathrm{KL}(p \,\|\, q) = \sum_{u} p(u) \log \frac{p(u)}{q(u)}$ is the Kullback--Leibler divergence, with the sum replaced by an integral for continuous $u$.
The metric quantifies the change in the task posterior caused by the semantic noise process, and it is task specific, since it is defined through the decoder $D_k$. Two stages with identical NMSE can contribute very differently to $\Delta_k$, and the same stage can contribute very differently to two clients with different tasks. We prefer the KL form because it requires no labels and can be estimated directly from unlabeled calibration data. For classification tasks with a one-hot clean posterior, \eqref{eq:semdist} reduces to the excess cross-entropy of the noisy model. Next, we derive a per-stage, per-client weighting of the energy budget that reflects this structure, yielding accuracy targets that are non-uniform across stages and clients.

\vspace{-0mm}\subsection{Sensitivity Decomposition}\label{sensitivity}
\vspace{-1mm}\begin{proposition}\label{prop:phi}
For small noise levels and independent noise across stages, the semantic distortion admits the decomposition
\vspace{-2mm}\begin{equation}\label{eq:sensitivity}
\Delta_k(\bm{\epsilon}_k) \approx \sum_{s \in \mathcal{S}} \phi_k^{(s)} \big(\epsilon_k^{(s)}\big)^2,
\vspace{-2mm}\end{equation}
where the \emph{semantic sensitivity} measures how strongly noise at stage $s$ degrades the task posterior of client $k$, per unit of NMSE, and is denoted by $\phi_k^{(s)}$ and given by
\vspace{-1mm}\begin{equation}\label{eq:phi}
\phi_k^{(s)} \triangleq \frac{\bar{S}^{(s)}}{2}\, \E_{\bx_k}\Big[\mathrm{tr}\Big(\bJ_k^{(s)}(\bx_k)^{\top}\,\bF_k(\bx_k)\,\bJ_k^{(s)}(\bx_k)\Big)\Big].
\end{equation}
\end{proposition}
Here $\bF_k = \mathrm{diag}(p_k) - p_k p_k^\top$ is the Fisher information matrix of the clean task posterior $p_k(u \mid \bx_k)=\mathrm{softmax}(\bm{\ell}_k)_u$ with respect to the decoder logits $\bm{\ell}_k$, defined as $\E_{u\sim p_k}\big[\nabla_{\bm{\ell}_k} \log p_k(u)\,\nabla_{\bm{\ell}_k} \log p_k(u)^\top\big]$. For a continuous task variable, $\bF_k$ is replaced by the Fisher matrix of the corresponding parametric posterior, and the derivation is unchanged. Here $\by_k^{(s)}$ denotes the noise-free signal at which the stage-$s$ noise is injected, namely the clean pre-activation $\by_k^{(\ell)}\triangleq\bW^{(\ell)}\bx_k^{(\ell)}$ for $s=\ell$ and the latent $\by_k^{(\mathrm{ch})}\triangleq\bz_k$ for $s=\mathrm{ch}$. The matrix $\bJ_k^{(s)}(\bx_k)\triangleq\partial\bm{\ell}_k/\partial\by_k^{(s)}$ is the Jacobian of the decoder logits with respect to this signal. We treat all signals as real vectors by stacking real and imaginary parts.

\begin{IEEEproof}[Proof sketch] Due to space limitations, we only sketch the derivation in three steps. First, let $\delta\bm{\ell} \triangleq \bm{\ell}(\bm{\epsilon}_k) - \bm{\ell}_0$ be the deviation of the decoder logits from their clean values caused by the noise. A second-order Taylor expansion of $\mathrm{KL}(p_k\,\|\,\widehat{p}_k)$ in $\delta\bm{\ell}$ around $\delta\bm{\ell}=\boldsymbol 0$ (where the KL divergence is zero) has a vanishing first-order term, since $\E_{p_k}[\nabla_{\bm{\ell}} \log p_k] = \sum_u \nabla_{\bm{\ell}} p_k(u) = \boldsymbol{0}$, and its second-order term is governed by the Fisher matrix, so
\vspace{-1mm}\begin{equation}
\mathrm{KL}(p_k \,\|\, \widehat{p}_k) = \tfrac{1}{2}\,\delta\bm{\ell}^\top \bF_k\,\delta\bm{\ell} + O(\|\delta\bm{\ell}\|^3).
\end{equation}
Second, a first-order Taylor expansion of the logits around the clean signals $\{\by_k^{(s)}\}$ gives $\delta\bm{\ell} = \sum_{s}\epsilon_k^{(s)}\,\bJ_k^{(s)}\bm{\nu}^{(s)} + O(\|\bm{\epsilon}_k\|^2)$, where $\epsilon_k^{(s)}\bm{\nu}^{(s)}$ is the noise injected at stage $s$. By the definition of the root NMSE, $\bm{\nu}^{(s)}$ is zero-mean with per-entry variance equal to the signal power $\bar{S}^{(s)}$, and independent across stages. The Jacobians are evaluated on the clean trajectory of the recursion \eqref{eq:recursion}, so corrections from noise propagated through subsequent stages enter at higher order. Third, averaging over the noise removes every cross-stage term, and each within-stage quadratic form becomes a trace via $\E[\bm{\nu}^\top\bM\bm{\nu}] = \bar{S}^{(s)}\mathrm{tr}(\bM)$ for any matrix $\mathbf{M}$, here $\mathbf{M}=\bJ_k^{(s)\top}\bF_k\bJ_k^{(s)}$.  Averaging over $\bx_k$ then yields \eqref{eq:sensitivity} and \eqref{eq:phi}. 
\end{IEEEproof}

 In practice the sensitivities are estimated offline without evaluating \eqref{eq:phi} directly, by injecting small single-stage noise and fitting the measured distortion to $\phi_k^{(s)}(\epsilon^{(s)})^2$ over a few probe amplitudes, and they do not depend on the channel realization. Although the encoder is shared, the sensitivities differ across clients because $\bF_k$ is induced by each client's decoder and task. For a coarse task most stages have low sensitivity, while a fine-grained task leads to high sensitivity at specific stages. The quantity $\phi_k^{\mathrm{ch}}$ can be interpreted as the semantic sensitivity of the communication link of client $k$, expressed in the same units as the sensitivities of the computation layers.

\vspace{-2mm}\subsection{Unified Water-Filling Allocation}\vspace{-2mm}
We formulate the allocation of the total client energy budget $\Gamma$ across clients and stages as the minimization of the network semantic distortion over the precisions $\gamma_k^{(s)}$. Expressing \eqref{eq:sensitivity} in terms of the precision of each stage, $\gamma_k^{(s)} = 1/(\epsilon_k^{(s)})^2$, the network distortion becomes a sum over all client-stage pairs of the sensitivity-weighted inverse precision, so that the semantic sensitivity weights the contribution of each stage. Formally, we write
\vspace{-3mm}\begin{equation}
\underset{\{\gamma_k^{(s)} > 0\}}{\text{minimize}} \; \sum_{k \in \cK} \sum_{s \in \mathcal{S}} \frac{\phi_k^{(s)}}{\gamma_k^{(s)}}
\quad \text{s.t.} \; \sum_{k \in \cK} \sum_{s \in \mathcal{S}} \omega_{k,s}\, \gamma_k^{(s)} \le \Gamma .
\label{eq:unified_wf}\vspace{-2mm}\end{equation}

\vspace{-1mm}\begin{lemma}\label{lem:wf}
Problem \eqref{eq:unified_wf} is convex, and its optimal solution, which we refer to as the unified semantic water-filling allocation, is, for every client-stage pair $(k,s) \in \cK \times \mathcal{S}$,
\vspace{-2mm}\begin{equation}\label{eq:wf}
\gamma_k^{(s)\star} = \frac{\Gamma \, \sqrt{\frac{\phi_k^{(s)}}{\omega_{k,s}}}}{\sum\limits_{(k',s') \in \cK \times \mathcal{S}} \sqrt{\phi_{k'}^{(s')} \omega_{k',s'}}}, \quad
\epsilon_k^{(s)\star} = \big(\gamma_k^{(s)\star}\big)^{-1/2}.
\vspace{-2mm}\end{equation}
\end{lemma}
\vspace{-4mm}\begin{proof}
The objective in \eqref{eq:unified_wf} is convex and the constraint is affine in the variables, and the constraint is active at optimality since the objective is decreasing in each variable. Stationarity of the Lagrangian gives $\phi_k^{(s)}/(\gamma_k^{(s)})^2 = \mu\, \omega_{k,s}$ for the multiplier $\mu > 0$, hence $\gamma_k^{(s)\star} \propto \sqrt{\frac{\phi_k^{(s)}}{\omega_{k,s}}}$. Enforcing the budget with equality yields \eqref{eq:wf}.
\vspace{-2mm}\end{proof}

Lemma~\ref{lem:wf} shows that the \emph{optimal precision $\gamma_k^{(s)\star}$ scales with the square root of the semantic sensitivity per unit of energy}, $\sqrt{\frac{\phi_k^{(s)}}{\omega_{k,s}}}$. Within a client, it assigns tighter NMSE targets to stages with higher sensitivity. Across computation and communication, it determines the split of the energy budget between the encoder layers and the uplink, which conventional designs fix in advance. Across clients, it assigns a larger share of the budget to clients with fine-grained tasks or weak uplinks. 
Designs that fix the link quality in advance, or that assign the same targets to all clients, are suboptimal whenever the ratios $\frac{\phi_k^{(s)}}{\omega_{k,s}}$ differ across client-stage pairs.

\vspace{-2mm}\section{Semantic Beamforming and Noise-Aware Training}\label{sec:training}\vspace{-2mm}
In this section, the energy allocation of Section~III is refined and converted into per-client beamforming gains, and the semantic encoder and decoder are trained jointly with the allocation.

\vspace{-3mm}\subsection{Allocation Refinement and Beamforming}\vspace{-1mm}
The expansion \eqref{eq:sensitivity} is accurate in the small-noise regime. For loose budgets, the split of the energy budget is refined by minimizing the exact network distortion $\sum_k \Delta_k$ over softmax-parameterized shares of $\Gamma$ for the $K(L+1)$ client-stage pairs, following \cite{yu2026analogrf}. The shares are initialized at the water-filling point \eqref{eq:wf} and updated by stochastic gradient descent with the Adam optimizer \cite{kingma2015adam}.

Given the resulting targets, the clients are coupled in the physical layer realization. The link targets are realized independently by the uplink power control $p_k = \sigma_{\mathrm{ul}}^2 \gamma_k^{\mathrm{ch}} / (|g_k^{\mathrm{ul}}|^2 \bar{S}_z)$ from \eqref{eq:chnmse}, subject to the client power limit. The computing targets, in contrast, must be delivered by a single broadcast beamformer per layer. By \eqref{eq:noisy_layer}, a layer-$\ell$ target $\epsilon_k^{(\ell)}$ requires the end-to-end gain $|a_k^{(\ell)}\beta_k^{(\ell)}| \ge \rho_k^{(\ell)} \triangleq \big(\kappa^{(\ell)}\big)^{-1/2}\big(\epsilon_k^{(\ell)}\big)^{-1}$, which costs the client the waveform energy $c_k^{(\ell)}\big(\rho_k^{(\ell)}\big)^2/|a_k^{(\ell)}|^2$. The BS therefore chooses the beamformer of each layer to minimize the total client waveform energy,  
\vspace{-3mm}\begin{equation}\label{eq:bf}
\begin{aligned}
&\underset{\mathbf f^{(\ell)}}{\mathrm{minimize}}\ \sum_{k\in\mathcal K}\frac{c_k^{(\ell)}(\rho_k^{(\ell)})^2}{|\mathbf h_k^H\mathbf f^{(\ell)}|^2} \\ 
 &\text{s.t.} \;\; \|\bff^{(\ell)}\|^2 \le P_{\mathrm{BS}}, \;\; |\bh_k^{H}\bff^{(\ell)}| \le a_{\max}\;\; \forall k,
\end{aligned}
\vspace{-2mm}\end{equation}
which is the per-layer problem of \cite{yu2026analogrf} with client weights determined by the semantic sensitivities of the tasks.  A client with high sensitivity at layer $\ell$ receives a larger beamforming gain at that layer, and a client with a coarse task receives a smaller gain. We refer to this design as semantic multi-user beamforming. Problem \eqref{eq:bf} is nonconvex because of the reciprocal terms in its objective, and is solved by the reduced-dimensional successive convex approximation of \cite{yu2026analogrf}. At each iteration, the gain $|\mathbf h_k^H\mathbf f|^2$ in the denominator is replaced by its first-order Taylor lower bound around the current iterate, which yields a convex surrogate, and $\mathbf f$ is restricted to the span of the channels, which reduces the dimension from $N_t$ to at most $K$. The semantic targets enter only through the weights $c_k^{(\ell)}(\rho_k^{(\ell)})^2$.

\vspace{-4mm}\subsection{Noise-Aware Training and Co-Design Loop}\label{sec:codesign}\vspace{-2mm}
The allocation above treats the semantic model as fixed. The model itself can be shaped to the noise it will encounter. We train the shared encoder and the task-specific decoders jointly, in multi-task fashion, with the full semantic noise process injected according to the current allocation,
\vspace{-2mm}\begin{equation}\label{eq:training}
\underset{E,\,\{D_k\}}{\text{minimize}} \;\; \sum_{k \in \cK} \E_{(\bx_k,u_k),\, \{\widetilde{\bn}\}}\Big[ \mathcal{L}_{\mathrm{task}}\big( \widehat{p}_k(u \mid \bx_k; \bm{\epsilon}_k), u_k \big) \Big],
\vspace{-2mm}\end{equation}
which extends deep JSCC training \cite{bourtsoulatze2019deep} from a single noisy bottleneck to a pipeline that is noisy at every stage and serves several tasks through one encoder.  Here, $u_k$ is the ground-truth task variable of client $k$ and $\mathcal{L}_{\mathrm{task}}$ is the task loss, for example the cross-entropy $-\log \widehat{p}_k(u_k \mid \bx_k; \bm{\epsilon}_k)$ for classification. Training and allocation are alternated as follows. (i) Train $E$ and $\{D_k\}$ under the current profiles $\{\epsilon_k^{(s)}\}$; (ii) re-estimate the sensitivities $\{\phi_k^{(s)}\}$ by the probe-noise fit of Section~III-B; (iii) update the allocation by \eqref{eq:wf}, and realize it by the beamforming problem \eqref{eq:bf} and the uplink power control. These steps repeat until the network distortion at the fixed budget stops decreasing.  

\vspace{-2mm}\begin{remark}
 Since the BS streams the encoder, a retrained model reaches all clients by changing the broadcast waveform, so the shared knowledge base stays synchronized by construction and the loop can track slow changes in task statistics, uplink quality, or energy budget.
\end{remark}

\vspace{-4mm}\section{Simulation Results}\vspace{-2mm}
\label{sec:sim_results}

We validate the sensitivity decomposition and the unified water-filling allocation of Lemma~\ref{lem:wf} on a multi-task SC network trained on MNIST.\footnote{Code is available at \url{https://github.com/christokt/semcom-analog-rf}.}

\subsubsection{Setup}
The shared encoder $E(\cdot;\{\bW^{(\ell)}\}_{\ell=1}^{L})$ has $L=3$ linear
layers ($784 \to 256 \to 64 \to 16$) with $\tanh$ activations on the first two
layers, giving a stage set $\mathcal{S} = \{\mathrm{L1}, \mathrm{L2},
\mathrm{L3}, \mathrm{ch}\}$ of size $S=4$ (three encoder layers plus the uplink
channel). $K=10$ clients observe MNIST digits, and half of them perform a fine-grained task, identifying the digit (10 classes), while the other half perform a coarse task, deciding whether the digit is at least 5 (2 classes). All clients share the same encoder, and each has its own decoder $D_k(\cdot)$.

Clients are heterogeneous in two respects. First, each client observes a non-i.i.d. local slice of the data, with per-client class proportions drawn from a $\mathrm{Dirichlet}(\alpha)$ distribution over the ten digit classes with $\alpha = 0.12$. Clients in the same task group therefore differ in their semantic sensitivity profiles $\bm\phi_k = (\phi_k^{(1)}, \ldots, \phi_k^{(S)})$ rather than sharing one group-level value. Second, each client
is assigned a hardware/channel quality factor (log-normal, $\sigma=0.6$) that
scales its per-stage energy coefficients $\omega_{k,s}$, on top of per-layer
jitter (log-normal, $\sigma=0.5$), modeling the fact that real analog
front-ends and uplink channels vary in quality well beyond mild noise around a
common baseline. Semantic sensitivities $\phi_k^{(s)}$ are estimated offline
via the probe-noise least-squares fit of Section~\ref{sensitivity}, using
probe levels $\epsilon_{\mathrm{probe}} \in \{0.03, 0.06, 0.10, 0.15\}$.

We compare three allocation schemes at each total energy budget $\Gamma$: (i)
\emph{uniform precision}, a common $\epsilon$ across all clients and stages;
(ii) \emph{client-blind water-filling}, which solves the water-filling
problem over stages only, using sensitivities and energy coefficients
aggregated across clients; and (iii) the \emph{proposed unified
water-filling} of \eqref{eq:unified_wf}, which allocates jointly over clients
and stages using the per-client $\phi_k^{(s)}$ and $\omega_{k,s}$.

\subsubsection{Sensitivity profile}
\begin{figure}[t]
    \centering
    \includegraphics[width=0.75\columnwidth]{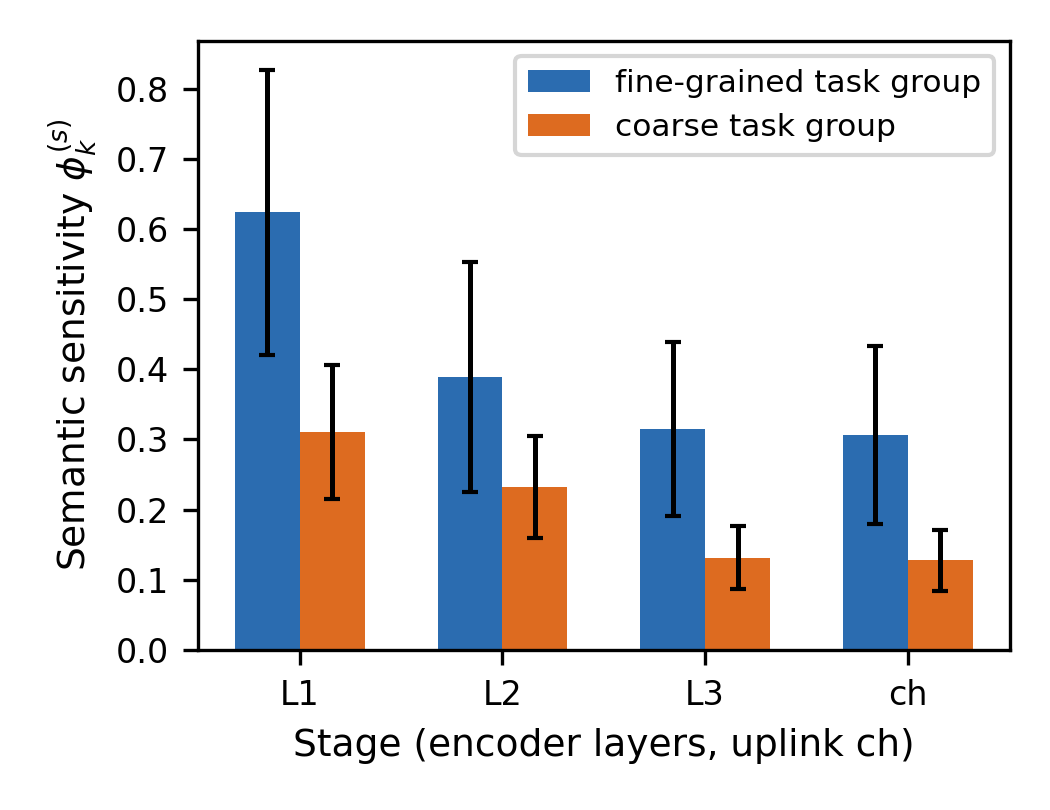}
  \vspace{-6mm}  \caption{Estimated semantic sensitivity $\phi_k^{(s)}$ by stage and task
    group, averaged over clients (error bars: one standard deviation across
    clients in the group).}
    \label{fig:sensitivity_profile}
\vspace{-5mm}\end{figure}

Fig.~\ref{fig:sensitivity_profile} shows the estimated $\phi_k^{(s)}$,
averaged within each task group with error bars showing the spread across
clients. It shows that sensitivity decreases
monotonically from the first encoder layer toward the channel stage, since
noise injected early propagates through more downstream nonlinearities, and
the fine-grained task group is uniformly more sensitive than the coarse task
group at every stage, since discriminating ten classes requires finer
distinctions in the latent representation than a binary decision. The
non-trivial error bars confirm that sensitivity also varies substantially
\emph{within} a task group due to the clients' differing local data
distributions, which is precisely the structure the unified allocation is
designed to exploit.

\subsubsection{Distortion-energy tradeoff}
\begin{figure}[t]
    \centering
    \includegraphics[width=1.01\columnwidth]{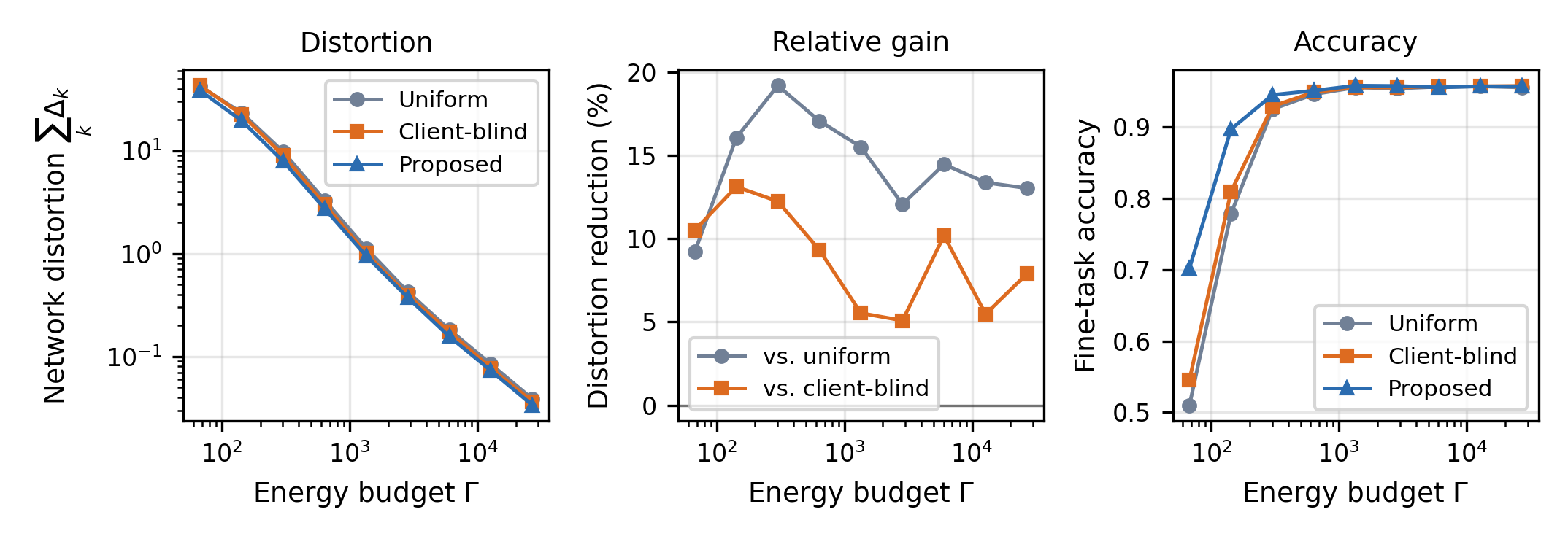}
   \vspace{-8mm} \caption{Distortion-energy tradeoff on MNIST. Left: network semantic
    distortion vs.\ total energy budget $\Gamma$ (log-log). Middle:
    distortion reduction of the proposed unified water-filling relative to
    uniform precision and client-blind water-filling. Right: fine-grained
    task accuracy vs.\ $\Gamma$.}
    \label{fig:distortion_tradeoff}
\vspace{-5mm}\end{figure}

Fig.~\ref{fig:distortion_tradeoff} sweeps the total client energy budget $\Gamma$ over a $26$~dB range, for each scheme,
the network semantic distortion $\sum_k \Delta_k$ (left panel), the
distortion reduction of the proposed scheme relative to each baseline
(middle panel), and fine-task accuracy (right panel). The network distortion
of all three schemes decays as a power law in $\Gamma$, consistent with the
$\Delta_k \propto 1/\gamma_{k,s}$ scaling of the second-order expansion. The
proposed unified water-filling dominates both baselines at every budget we
tested, reducing network distortion by $9.2\%$ to $19.2\%$ relative to
uniform precision and by $5.1\%$ to $13.1\%$ relative to client-blind
water-filling. In the energy-starved regime ($\Gamma \approx 67.5$), fine-task accuracy is
$0.702$ under the proposed scheme versus $0.546$ for client-blind and $0.510$
for uniform allocation, a gap that closes once the budget is large enough
that all three schemes can afford low noise everywhere. This matches the
intuition behind Lemma~\ref{lem:wf} that when energy is scarce, correctly
identifying \emph{which} client-stage pairs are most sensitive and least
costly to protect matters most. When energy is abundant, every scheme can
simply drive $\epsilon_{k}^{(s)} \to 0$.

\subsubsection{Noise-aware co-design}
\begin{figure}[t]
    \centering
    \includegraphics[width=0.81\columnwidth]{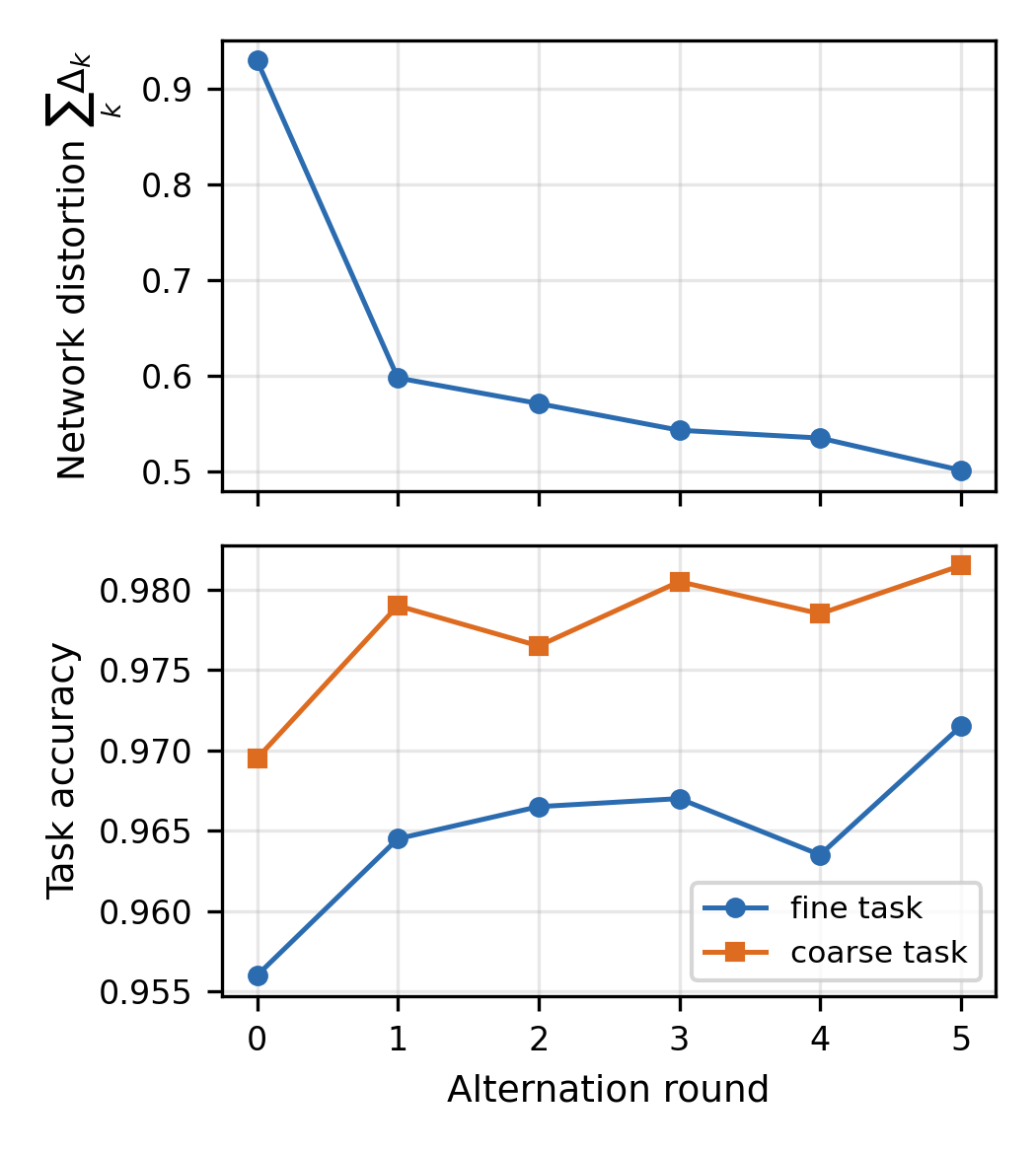}
  \vspace{-4mm}   \vspace{-3mm}\caption{Noise-aware co-design loop at fixed budget $\Gamma \!\approx\!
    1350$: network semantic distortion (left) and per-task accuracy (right)
    across alternation rounds.}
    \label{fig:codesign}
\vspace{-6mm}\end{figure}

Fig.~\ref{fig:codesign} shows the alternating co-design loop of
Section~\ref{sec:codesign}, which alternates retraining the shared encoder
and decoders under the currently allocated noise profile with re-solving the
unified water-filling problem, at a fixed budget $\Gamma \approx 1350$, which corresponds to $13$ dB.
Network distortion falls from $0.930$ at round $0$ (the clean-trained model,
evaluated under noise for the first time) to $0.501$ after five rounds, a
$46.1\%$ reduction, while fine-task accuracy improves from $0.956$ to
$0.972$ and coarse-task accuracy from $0.970$ to $0.982$. Most of the gain is
realized within the first round, indicating that a single retraining pass
under the allocated noise profile already captures most of the benefit of
noise-aware training.

\vspace{-2mm}\section{Conclusion}\vspace{-2mm}
This paper proposed an end-to-end SC architecture in which the semantic encoder is computed by the radio itself through analog RF computing. We unified analog computing noise and channel noise as a single semantic noise process, defined a semantic distortion metric on the task posterior, and derived a closed-form water-filling allocation of one energy budget jointly over all clients, encoder layers, and uplinks, whose realization yields a beamforming problem with semantically determined per-client gain requirements. On MNIST at matched total client energy, the unified allocation reduced the network semantic distortion by 9.2\%--19.2\% relative to uniform precision and by 5.1\%--13.1\% relative to a client-blind allocation, and noise-aware co-design reduced it by a further 46\%. Future work includes transformer-based semantic models with input-dependent MVMs, analog computation of the decoder at the BS, multi-user semantic multiple access over the uplink, and combining analog RF computing with over-the-air computation for multi-agent semantic fusion.
\vspace{-2mm}\bibliographystyle{IEEEbib}
\def\baselinestretch{0.84}
\bibliography{refs}

\end{document}